\documentclass[epj]{svjour}
\usepackage{epsfig, amsmath}
\begin{document}
\title{Signatures of Mach shocks at RHIC}
\author{Thorsten Renk\inst{1} \inst{2}
}                     
\offprints{}          
\institute{Department of Physics, PO Box 35 FIN-40014 University of Jyv\"askyl\"a, Finland \and Helsinki Institut of Physics, PO Box 64 FIN-00014, University of Helsinki, Finland}
\date{Received: date / Revised version: date}
%
\abstract{
Hard partons propagating through hot and dense matter loose energy, leading to the observed depletion of hard hadron spectra in nucleus nucleus collision as compared to scaled proton proton collisions. This lost energy has to reappear in the medium due to the conservation of energy. Apart from heating the medium, there is the possibility that a propagating collective mode is excited. We outline a formalism that can be used to track the propagation of such a mode through the evolving medium if its dispersion relation is known and to calculate the resulting distortion of hadronic spectra at freeze-out. Using this formalism, we demonstrate within a detailed picture of the evolution of the expanding system and the experimental trigger conditions that the assumption of a sound mode being excited is in line with 2-particle correlation measurements by  PHENIX and STAR for a semi-hard trigger. In this case, the measurement is sensitive to the averaged speed of sound in the hot matter. We also make suggestions how this sensitivity can be improved.
\PACS{
      {25.75.-q}{Relativistic heavy-ion collisions}   \and
      {25.75.Gz}{Particle correlations}
     } 
} 
\maketitle
\section{Introduction}
\label{intro}

Energy loss of a high $p_T$ 'hard' parton travelling through low $p_T$ 'soft' matter has long been 
recognized as a promising tool to study the initial high-density phases of ultrarelativistic
hevay-ion collisions (URHIC) \cite{Jet1,Jet2,Jet3,Jet4,Jet5,Jet6}. However, if one considers the whole dynamical system created in the collision of two relativistic nuclei and not only the partons emerging from a particular hard scattering vertex, energy is not lost but rather redistributed into the medium. 

Measurements of angular correlations of hadrons associated with a given hard trigger allow, as a function of associate hadron momentum, to study how and at what scales this redistribution of energy and momentum takes place. Such measurements for semi-hard hadrons with 1 GeV $< p_T < $ 2.5 GeV associated with a trigger 2.5 GeV $< p_T <$ 4.0 GeV have shown a surprising splitting of the away side peak for all centralities but peripheral collisions, qualitatively very different from a broadened away side peak observed in p-p or d-Au collisions \cite{PHENIX-2pc}. For a harder trigger 2.0 GeV $ < p_T < $ 4 GeV and different associate hadron momentum cuts, a persistent large angle signal has been observed \cite{STAR-2pc}.

As most promising explanation for these findings, the assumption that Mach shockwaves are excited by the energy lost from the hard parton to the medium has been brought forward \cite{Shuryak,Stoecker}. In the following, we investigate under what conditions signals from such shockwaves remain observable in the dynamical environment of a heavy-ion collision, provided that a realistic description of the evolving medium and the experimental trigger conditions are taken into account. Here, we summarize and expand on \cite{Mach1} and \cite{Mach2}.

\section{Shockwave excitation, propagation and freeze-out}

For the time being we focus on central collisions only. For the description of the evolving medium, we employ a parametrized evolution model \cite{RenkSpectraHBT,Synopsis} which is known to describe bulk matter spectra and HBT correlations well.
 Since we're interested in the deposition of lost jet energy
into the medium, our first task is to determine the spacetime pattern of energy loss. We use the quenching weights from
\cite{QuenchingWeights} to obtain the probability $P(\Delta E)$ to lose the amount of energy $\Delta E$ from the two
key quantities plasma frequency

\begin{equation}
\omega_c({\bf r_0}, \phi) = \int_0^\tau d \xi \xi \hat{q}(\xi)
\end{equation}
and averaged momentum transfer
\begin{equation}
(\hat{q}L) ({\bf r_0}, \phi) = \int_0^\tau d \xi \hat{q}(\xi)
\end{equation}

in a static equivalent scenario which are calculated along the path of the hard parton through the medium. 
Since we are not interested in folding the result with a steeply falling spectrum but rather into the
energy deposited on average in a given volume element we focus on the average energy loss $\langle \Delta E \rangle =
\int_0^\infty P(\Delta E) \Delta E d\Delta E$ in the following.

We make no attempt to calculate the microscopic mechanism by which a propagating mode is excited. 
In \cite{Shuryak} it was shown that only a certain class of source terms leads to a propagating sound 
wave in a hydrodynamical medium. In particular source terms which directly couple energy and momentum 
lost by the hard parton into the local fluid cell do not lead to angular correlations as seen by experiment.
This is in fact consistent with the findings of \cite{Uli} where no such correlations were found in a 2d Bjorken hydro code 
under the assumption of a direct coupling. A possible microscopical excitation mechanism might be colour wakes induced in the medium \cite{Wake}.

For the purpose of the present work, we assume that a fraction $f$ of the energy and momentum lost to the medium 
excites a shockwave characterized by a dispersion relation $E = c_s^2 p$ and a fraction
$(1-f)$ in essence heats the medium and leads to some amount of collective drift along the jet
axis to conserve longitudinal momentum.

We calculate the speed of sound $c_s$ locally from a quasiparticle description of the equation of state as
measured on the lattice \cite{STW} as $c_s = \partial p(T)/ \partial \epsilon(T)$.
This EOS shows a significant reduction of $c_s$  as one approaches the phase transition but doesn't lead to a mixed phase. The dispersion relation along with the energy and momentum deposition determines the initial angle of propagation
of the shock front with the jet axis (the 'Mach angle') as $\phi = \arccos c_s$. We discretize the time into small intervals $\Delta \tau$, calculate the energy deposited in that time as
$E(\tau) = \Delta \tau \cdot dE/d\tau$ we then propagate the part of the shockfront remaining in the midrapidity
slice (i.e. in the detector acceptance). Each piece of the front is propagated with the local speed of sound and the angle of propagation is
constantly corrected as 
\begin{equation}
\label{E-c_s}
\phi = \arccos \frac{\int_{\tau_E}^\tau c_s(\tau) d\tau }{(\tau - \tau_E)}
\end{equation}
where $c_s(\tau)$ is determined by the propagation path.

Once an element of the wavefront reaches the freeze-out condition $T = T_F$, a hydrodynamical
mode cannot propagate further. We assume that this point that the energy contained in the shockwave
is not used to produce hadrons but rather is converted into kinetic energy. In the local
restframe, we then have a matching condition for the dispersion relations
\begin{equation}
E = c_s^2 p \quad \text{and} \quad E = \sqrt{M^2 + p^2} - M
\end{equation} 
where $M = V \left(p(T_F) + \epsilon(T_F)\right)$ is the 'mass' of a volume element at freeze-out temperature.

Once we have calculated the additional boost $u_\mu^{shock}$ a volume element receives from the shockwave using the
matching conditions, we use the Cooper-Frye formula
\begin{equation}
\label{E-CF}
E \frac{d^3N}{d^3p} =\frac{g}{(2\pi)^3} \int d\sigma_\mu p^\mu
\exp\left[\frac{p^\mu (u_\mu^{flow} + u_\mu^{shock}) - \mu_i}{T_f}\right]
\end{equation}
to convert the fluid element into a hadronic distribution.
The resulting momentum spectrum is thus a thermal two component spectrum resulting from an integrations involving
volume not part of the shockwave and volume receiving an additional boost from the shockwave.

\section{Simulation of the trigger conditions}

We simulate the 
PHENIX trigger conditions as closely as possible using a Monte-Carlo approach. 
We start by generating vertices with a distribution weighted by the nuclear overlap
\begin{equation}
T_{AA}({\bf b}) = \int dz \rho^2({\bf b},z).
\end{equation}

We then determine the jet momentum and parton type by randomly sampling partonic transverse momentum spectra generated
by the VNI/BMS parton cascade as described in \cite{VNI-LPM}. Calculating the energy
loss of the near side parton, we decide if the experimental trigger condition is fulfilled.
Since the experiment triggers on a hard hadron in the transition between the recombination and fragmentation regime, 
the model at this point cannot implement the trigger condition exactly. Instead, we require the trigger
condition to be fulfilled by the parton and have checked that the model results do not
change significantly when the near side trigger threshold is increased by 2 GeV.
We note that this procedure places the vertices fulfilling the trigger condition close to the surface 
of the produced matter, i.e. in our model the
medium is rather opaque, in agreement with the conclusions of \cite{Fragility,Dijets}.

Once the vertex and momentum of a near side jet has passed the trigger condition, we 
determine the direction of the away side parton in the transverse plane and rapidity. In order to
take into account intrinsic $k_T$, we do not propagate the away side directly opposite to the near side jet but allow for a random angle. We have verified that this distribution, folded with the width of the near
side peak reproduces the width of the far side peak in the case of d-Au and 60-90\% peripheral Au-Au collisions.

With vertex, energy and direction of the away side jet fixed, we calculate $dE/d\tau$ of the outgoing parton.
We stop the calculation when a significant fraction of the energy is lost to the medium. In each event we assume 
that a fraction $(1-f)$ of the energy lost from the hard parton heats the medium.
We assume that due to momentum conservation this contribution not leading to shockwaves will lead to additional flow into the direction of the
original away side parton. Likewise, we account for the possibility of a punchthrough if the initial 
vertex is very peripheral and  both near and away side parton propagate near-tangential to the 
surface.

\section{The role of transverse flow}

It has been argued that the includion of transverse flow leads to a substantial distortion of the Mach cones \cite{Satarov}. We observe this effect when we consider the spatial position of the cone at a given time \cite{Mach1}. However, as apparent from Eq.~(\ref{E-CF}), the measured effect of the shockwave is not manifest in position space but rather (via a boost to volume elements undergoing freeze-out) in momentum space.

When the associate hadron cut is set well above the bulk matter momentum scales at freeze-out $p \approx 3 T_F \approx 400$ MeV, the observation takes place in the tail of the boosted thermal distribution. Under these conditions, the yield is very sensitive to boosts and almost all observed yield comes from a region where the transverse flow is aligned with the emitted particle momentum. In the presence of a shockwave, the signal at 1 GeV momentum is maximal if $u_\mu^{shock}$ and $u_\mu^{flow}$ are parallel (in fact, the observed yield for $u_\mu^{shock} \parallel u_\mu^{flow}$ is about 9 times larger than the yield for configurations in which $u_\mu^{shock} \perp u_\mu^{flow}$). However, if flow velocity and shock propagation are (approximately) aligned, no distortion of the Mach angle can take place. Thus, for an associate momentum cut beyond typical thermal scales, only configurations in which the flow does not distort the Mach angle are visible. 

\begin{figure*}[!htb]
\epsfig{file=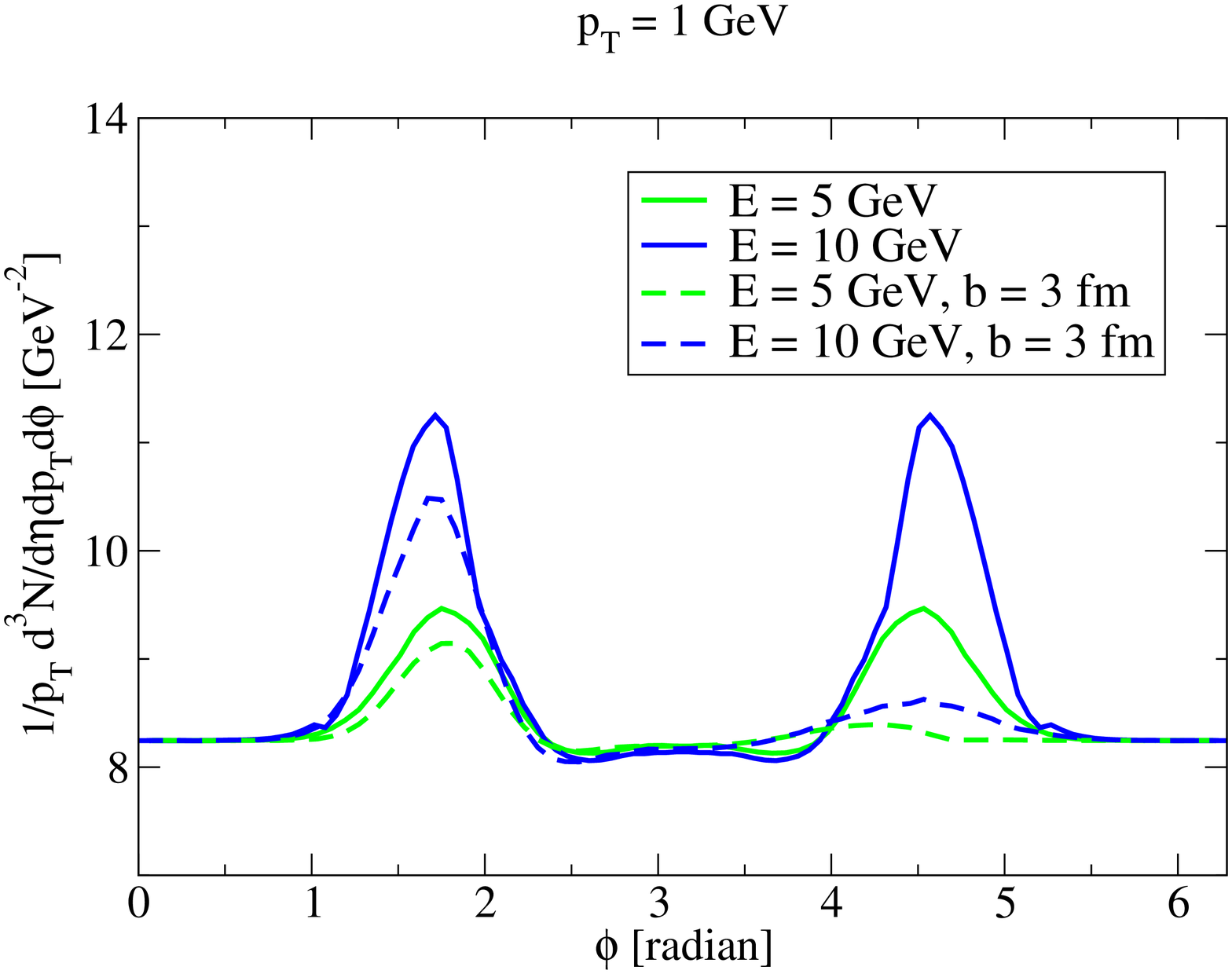, width=7.5cm}\epsfig{file=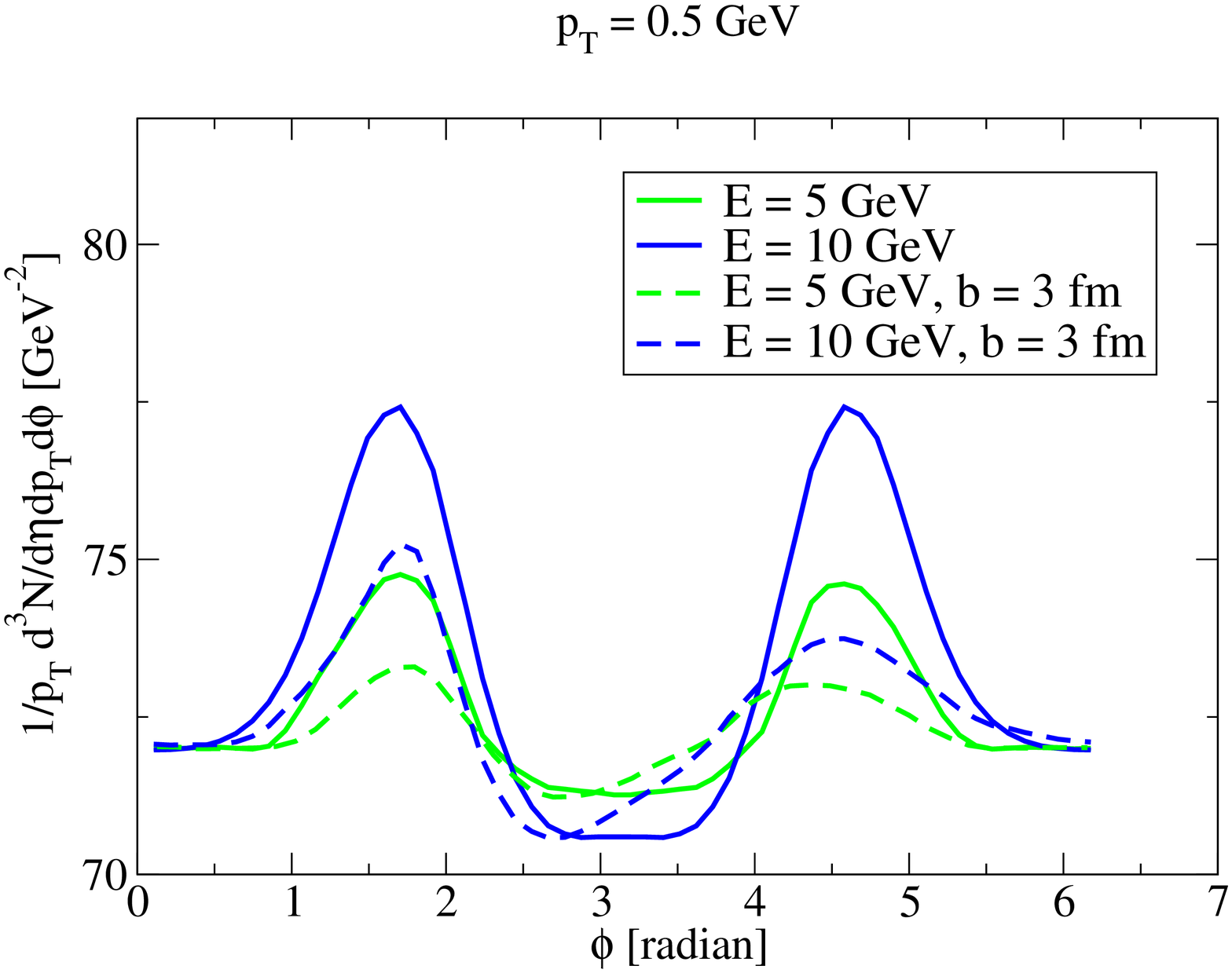, width=7.5cm}
\caption{
\label{F-Disappearance}The effect of transverse flow on the observable correlation peak for configurations in which transverse flow 
and cone
are aligned (solid) and orthogonal (dashed, $b=3$ fm) for associate hadron $p_T = 1.0$ GeV 
(left panel) and $500$ MeV (right panel). The apparent disappearance of the correlation signal 
at high $p_T$ when flow and shockwave are not aligned is clearly visible. At lower $p_T$ the 
signal reappears significantly broadened, reflecting the distortion of the Mach angle by transverse 
flow. All calculations here are for a single event with given parton energy $E$ and assuming no 
intrinsic $k_T$ and $y=0$ as the position of the away side parton. Correlation strength in the 
direction of the away side parton have been suppressed for clarity. }
\end{figure*}

On the other hand, momentum conservation dictates that the correlation signal cannot simply vanish for configurations in which shock and flow are not aligned. In this case, a broader peak structure is recovered with a lower $p_T$ cut. This is illustrated in Fig.~\ref{F-Disappearance}.

\section{The role of longitudinal flow}

\label{S-Rapidity}

Even if the trigger parton is confined to be at midrapidity, the rapidity of the away side parton is not known. The differential production cross-section of hard partons in A-A collisions can be obtained in leading order 
pQCD by folding the two particle production cross-sections $d\hat{\sigma}/d\hat{t}$ with the nuclear parton 
distribution functions $f_{i,j/A}$ (here we use \cite{NPDF}, $Q^2$ dependence has been suppressed in the following expressions):
\begin{eqnarray}
\label{cross-y}
\frac{d^3\sigma^{AA \rightarrow kl+X}}{dp_T^2dy_1dy_2}= \negthickspace \sum_{i,j} x_1 f_{i/A}(x_1)x_2 f_{j/A}(x_2) \frac{d\hat{\sigma}^{ij\rightarrow kl}}{d\hat{t}}
 \end{eqnarray}

If the outgoing partons are at rapidities $y_1$ and $y_2$, $x_1$ and $x_2$
are determined by:

\begin{equation}
x_{1,2} = \frac{p_T}{\sqrt{s}} \left[\exp(\pm y_1) + \exp(\pm y_2)\right] 
\end{equation}

The conditional probability distributions $P_{}(y)$ of producing an away-side parton  at rapidity $y$ can then be calculated from the normalized cross-section (Eq.~(\ref{cross-y})) given the trigger parton at $y_1=0$. For the dominant processes, it is a rather wide plateau between $-2 < y < 2$. 

This raises an obvious question: Since the Mach cone is symmetric around the away side parton direction, a measured large angle correlation near midrapidity also implies large rapidity correlations at zero transverse angle.  Thus, the averaging over $P(y)$ will tend to smear the signal measured at midrapidity out towards smaller angles as compared to a simple midrapidity projection of teh cone for an away side parton emerging at $y=0$.

However, since the shock wave travels with $c_s$ in the local rest frame, the spatial position the of the shock front has to be determined by solving the characteristic
equation:
\begin{eqnarray}
\left.\frac{dz}{dt}\right|_{z=z(t)}=\left.\frac{u(z,R,t)+c_s(T(z,R,T))}{1+u(z,R,t)c_s(T(z,R,t))}\right|_{z=z(t)}.
\end{eqnarray}

\begin{figure}[htb]
\epsfig{file=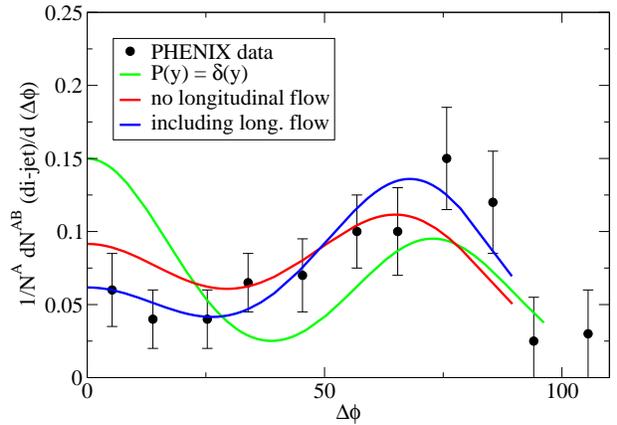, width=8cm}
\caption{\label{F-Elongation}Calculated 2-particle correlation under the assumption that a) the away side parton is always at midrapidity and the excited mode doesn't couple to flow (green) b) using realistic $P(y)$ and assuming that the excited mode doesn't couple to flow (red) and c) including realistic $P(y)$ and longitudinal flow elongation. }
\end{figure}

We use this equation to infer the longitudinal boost for an element of the Mach cone at forward rapidity. In essence, this means that a Mach cone in $\phi,y$-space is elongated significantly in $y$ direction by longitudinal flow because a sound wave (unlike emitted particles) propagates in rapidity space even in a Bjorken expansion. For a Bjorken expansion this elongation amounts to a about $1.5$  units rapidity. The net effect of this elongation is that the Mach signal, i.e. the angular correlation at given rapidity are much less sensitive to the rapidity averaging procedure than naively expected. We illustrate this in Fig.~\ref{F-Elongation}.

\section{The dip}

It is sometimes argued that it is the dip at the expected position of the away side parton (at 0 degrees in Fig.~\ref{F-Elongation}) along with the rise of the correlation strength at some large angle with respect to this which constitutes the signal of Mach shocks. However, even in our analysis, the fraction of energy exciting the shockwave is only 75\%, thus 25\% of the energy lost by the away side parton ends up associated with its original direction. It is easily apparent from Fig.~3 in \cite{Mach1} that the appearance of the dip is strongly linked to the fact that the {\em dominant} fraction of lost energy excites a shockwave, not to the fact that a shockwave is present as such. It is equally instructive to study Fig.~\ref{F-Elongation} in the present work. The same signal that would be interpreted as having a dip if the rapidity averaging is carried out fully would change dramatically if the position of the away side parton could be confined to midrapidity. The reason is that while in the rapidity-averaged situation the away side parton is outside the narrow acceptance region around midrapidity in most events (and hence only a part of the wider ring ends up within the acceptance), for specified away side parton rapidity suddenly all correlation strength associated with the original direction ends up always inside the acceptance region whereas the Mach ring is still for the most part outside.

Thus, the presence or absence not only hinges on the question if the cone is the dominant mode excited by lost energy but also on the experimental cuts. It is also expected that for increasing trigger energy a larger fraction of original away side partons emerges from the medium, subsequently undergoing vacuum fragmentation. Such a contribution would likewise fill the dip region.

A more reliable characteristic of the Mach cone is that the angle to which the away side correlation extends is a property of the medium (it is solely determined by the path-averaged speed of sound, cf. Eq.~(\ref{E-c_s}) and independent of both trigger energy and associate hadron cut. Such a behaviour is difficult to obtain in radiative processes from the away side parton where it is expected that for increasing $k_T$ of emitted secondaries (and hence increasing associate hadron cut) the angle shrinks. A signal qualitative in agreement with this expectation of unchanged angle is observed by the STAR collaboration \cite{STAR-2pc}.

\section{Precision determination of the Mach angle}

If the observed correlation pattern is indeed caused by Mach shocks, a determination of the precise Mach angle offers an opportunity to probe the equation of state of the underlying system. In \cite{Mach1} we have shown that there is some sensitivity to the averaged speed of sound - the difference between assuming a phase transition with a soft point (with small $c_s$) and assuming $c_s \approx 0.55$ at all times amounts to a shift of some 18 degrees in peak position. However, more realistic variations of $\langle c_s \rangle$ accessible by experiment (e.g. by a variation of initial temperature for different $\sqrt{s}$ or a variation of decoupling temperature by going to different system size) to confirm our ideas about the medium EOS require a determination of the peak position by a precision of the order of a few degrees ($O(1-5^\circ)$). This is beyond the ability of the current data and made difficult by the fact that in the present measurement the peak is always broadened.

In a static, homogeneous cold medium, the correlation signal would closely resemble a $\delta$-function at $\phi_{Mach}$ (which is also different from $\phi_{Mach}$ in the dynamical evolution). In the dynamical model, the correlation signal appears broadened for a number of reasons: 1) Different elements of the cone are excited at different times and probe different parts of the (inhomogeneous) medium. This is dispersion introduced by the spacetime structure of the medium. 2) The intrinsic $k_T$ randomly influences the angular position of the away side parton. Thus, in averaging over many events all peaks get smeared with the induced angular spread. 3) The thermal motion of hadrons at freeze-out leads to a thermal smearing of the correlation signal and 4) The rapidity averaging outlined in section \ref{S-Rapidity} leads to a widening and systematic shift towards small angles when averaging over many events is done.

However, some of these effects can be addressed by experiment: By requiring a hard dihadron as the trigger, i.e. a near side hadron and a (softer) away side hadron, the rapidity and intrinsic $k_T$-kick of the away side parton is known for each event and the Mach correlation can be studied relative to the measured away side parton direction. In doing so, 2) and 4) are eliminated as mechanisms leading to a widening of the peak. The thermal width can be somewhat reduced by increasing the momentum threshold of analyzed associate hadrons (thermal motion transverse to the associate hadron direction is of less relative importance if the associate hadron momentum is increased). However, one can not push this to arbitrarily high momenta as the presence of a shockwave signal requires that a fluid medium dominates the hadronic spectra (which ceases to be true above 2-3 GeV). The resulting peak in the correlation signal could be significantly narrower than what is observed now (cf. Fig.~\ref{F-Disappearance} where intrinsic $k_T$ broadening and rapidity averaging are absent since the plot shows a single event only).

Finally, measuring correlations for hadrons with a different freeze-out systematics (e.g. $\phi$ and $\Omega$ for which blast wave fits and their small scattering cross suggest early decoupling) could potentially offer a window to study both the systematic change in thermal width of the signal and a different Mach angle (since Eq.~(\ref{E-c_s}) is evaluated for a different upper limit).

\section{Conclusions}

While we have not presented a complete microscopical theory of the excitation of Mach shockwaves by hard partons travelling through a hot and dense medium, we have shown that if one makes the assumption that such shockwaves are excited, the resulting correlation pattern is consistent with the measured data even if a realistic trigger simulation and effects of the medium evolution are taken into account.

In particular, we have presented the (somewhat counterintuitive) finding that above a certain threshold in transverse momentum, the correlation signal is not significantly distorted by transverse flow. The underlying reason is peculiar for a hydro phenomenon --- since scales $O(\text{1 GeV})$ where the correlation has been measured by PHENIX are above the typical hydro scales, only alignment of flow and shockwave direction (and hence no distortion) can boost the hydro medium enough to create a signal.

We presented a similar finding within a discussion of the longitudinal dynamics. There, the elongation of the Mach cone by longitudinal flow, a phenomenon only taking place for a perturbation moving with a given speed {\em relative to the expanding medium}, proved to be a crucial ingredient in the survival of the signal after averaging over the (a priori unknown) position of the away side parton in rapidity.

While a detailed discussion of the excitation function as a function of trigger energy and associate cut still has not been done (the chief obstacle being a clean modelling of the transition between recombination and fragmentation as a mechanism for hadronization) we note that the absence of cone-like structures in high transverse momentum angular correlation measurements \cite{Dijets1,Dijets2} and their explanation in terms of hard partons emerging from the medium \cite{Dijets} are well in line with the observation that hydrodynamical modes are irrelevant at $p_T > 4$ GeV at RHIC.

In summary, while the current data certainly do not allow Mach shocks as the only explanation for the measured correlation pattern, it can be shown that they are a consistent explanation, and their nature as a hydrodynamical mode explains some characteristics found in the data in a natural way. A similarly conclusive case has yet to be made for alternative explantions for the correlation pattern seen in the data.

I would like to thank J.~Ruppert, V.~Ruuskanen, K.~Eskola, B.~M\"{u}ller, S.~Bass, R.~Fries,  P.~Jacobs and J.~Rak for valuable comments and discussions. This work was financially supported by the Academy of Finland, Project 206024. 

%
%

\end{document}